\documentclass[referee]{aa} 
\newcommand{\ltapprox}{\raisebox{-0.5ex}{$\,\stackrel{<}{\scriptstyle\sim}\,$}}

\begin{document}
\input psfig.tex
 
\title{$K$--band luminosity (mass) segregation in AC\,118 at $z=0.31$\footnote{Based on observations
collected at the European Southern Observatory, Chile, ESO N° 62.O-0369,
63.O-0115, 64.O-0236}
}
\author{S. Andreon}
\institute{INAF--Osservatorio Astronomico di
Capodimonte, Via Moiariello 16, 80131 Napoli, Italy\\
E-mail: andreon@brera.mi.astro.it}

\titlerunning{$K$--band luminosity (mass) segregation in AC\,118}

\date{Accepted 13 November 2001}
 
\abstract{ Using new observations of the galaxy cluster AC\,118 at
intermediate redshift ($z=0.31$) in the $K_s$ band, we were able to detect
the cluster from the center to half the Abell radius (1.5 Mpc, $H_0=50$ km
s$^{-1}$ Mpc$^{-1}$) and possibly to 2.0 Mpc. The analysis of both the
spatial distribution of galaxies of various luminosities and of the
luminosity function (LF) of galaxies in different cluster locations
strongly confirms and extends to larger clustercentric radii the
luminosity segregation found in a previous analysis of this cluster
restricted to a smaller cluster area: there is an excess of bright
galaxies in the cluster core (inside 250 Kpc) or a deficit of dwarfs in
the remain part of the cluster. Outside the cluster core and as far as 1.5
or even 2 Mpc, the giant--to--dwarf ratio is constant. Because of the
luminosity segregation, the LF of the AC\,118 shows a larger number of
bright galaxies per unit dwarf in the core than in other cluster
locations. All non--core LFs, computed at several cluster locations, are
compatible each other. These results hold both including or excluding the
galaxies located in an overdensity found in the far South of AC\,118 and in
the second clump in galaxy density at the cluster North--West. Since the
near--infrared emission is a good tracer of the stellar mass, we interpret
the segregation found as a mass segregation.  
 }
\maketitle 

\section{Introduction}

Field galaxies should experience interactions with the hostile cluster
environment during infall in the cluster because of harassment (Moore et
al. 1996), tidal tails and eventually mergers (Toomre \& Toomre 1972) and
possibly other cluster--specific phenomena. These effects influence
infalling galaxies well before they reach the cluster core.  Therefore, it
is important to observe galaxies when interactions are occurring, i.e. at
large clustercentric distances. The dependence of the galaxy properties on
clustercentric distance is therefore informative of cluster--related
phenomena. 

Studies of galaxy evolution in clusters often sample only the cluster core
because the cluster outskirts have a low contrast with respect to the
background galaxies, making the measure of cluster galaxy properties
subjected to large errors. Furthermore, the small field of view of the
available imagers, in particular in the near--infrared, makes the sampling
of the cluster outskirts expensive in telescope time. For these reasons,
cluster outskirts are less frequently studied, in particular in the
near--infrared, even if the near--infrared is very informative: for
example it is a good tracer of the stellar mass (Bruzual \& Charlot 1993;
Pierini, Gavazzi \& Boselli, 1996) and it is not too affected by
short star bursts (Bruzual \& Charlot 1993).  Figure 10 in Andreon (2001)
shows that there are only three investigations (de Propris et al. 1998;
Andreon \& Pell\'o 2000; Andreon 2001) sampling galaxies fainter than
$M^*+1$ and exploring radii larger than 0.4 Mpc. Since then, another work
has appeared (Tustin, Geller, Kenyon \& Diaferio 2001). The situation is now
rapidly changing thanks to 2MASS (Skrutskie et al. 1997) and large
panoramic infrared receivers, such as CIRSI (Mackay et al. 2000). 

In this paper we make use of a large panoramic near--infrared receiver, an
Hawaii chip, by imaging the intermediate redshift cluster AC\,118. AC\,118 has
been observed with three pointings, two of which image the cluster outskirts
and are presented in \S 2. The central pointing was, instead, 
presented in Andreon (2001, hereafter Paper I). In \S 3 the spatial
distribution of galaxies of various luminosities, the dwarf to giant radial
profile and the luminosity function at various cluster locations are
presented. In \S 4 we present a summary and discuss the results.

In this paper we assume $H_0=50$ km s$^{-1}$ Mpc$^{-1}$ and $q_0=0.5$. 
The choice of the cosmology has a small or null impact on the results 
because all the compared galaxies are at the same redshift (see Sect 4). 

\section{Data and data reduction}

AC\,118, also known as Abell 2744 ($\alpha , \delta = 00 \ 14 \ 19.5 \ -30
\ 23 \ 19, \ J2000$), is a cluster of galaxies at intermediate redshift
($z=0.3$, see Paper I for a summary of its properties). Its central region
has been imaged in the near--infrared $K_s$ band with SOFI at NTT (Paper
I). AC\,118S and AC\,118N, the northern and southern pointings of AC\,118,
have been observed in the $K_s$ band in September 18 and October 31, 1999,
respectively, with SOFI at NTT in the frame of an observational program
aimed at deriving the Fundamental Plane at $z\sim0.3$, as a complement to
our central pointing. SOFI is equipped with a $1024\times1024$ pixel
Rockwell ``Hawaii" array, with a 0.292 arcsec pixel size and a $5\times5$
arcmin field of view. The two pointings are offset, with respect to the
previous central pointing, by almost one SOFI field of view, to maximize
the survey area while keeping enough overlap to check the consistency of
the photometry of the three pointings (see Figure 1 for the pointing
layout).

Table 1 gives a summary of the characteristics of the data used in this paper.
Exposures times are shorter and observations are shallower, with respect
to the central field, by a factor $\sim3$ because of reduced time for
near--infrared observations.

The data have been reduced as described in Paper I. Briefly, images are
flat--fielded by means of differential dome flats and calibrated by means of
Persson et al. (1998) standard stars. The background is removed by a temporal
filtering of the images, using Eclipse (Devillard, 1997). The combining of the
individual exposures is performed by means of {\it imcombine} under IRAF,
making full use of the bad pixel mask and weights and aligning images without
resampling. As in Paper I, the dependence of the atmospheric absorption on
airmass is computed from the science data, since the target is
observed at different hour angles.

Two (minor) differences apply with respect to the data reduction described in
Paper I: a) the illumination correction is found to be significant, and applied
to, AC\,118S frames; b) a residual shallow spatial gradient is present in the
frames, even after the background subtraction performed via a filtering in
the time domain. Therefore, we introduce a further step in the background
subtraction, by fitting, and removing a plane to the background.

The two nights were photometric, as determined from the scatter of the zero
point of the standard stars and from the scatter, from frame to frame, of the 
instrumental magnitudes of a reference galaxy in the field of view. 
For objects in common between fields, that are independently calibrated, we
found systematic differences less than 0.01 mag, confirming the quality of the
observing nights and of the data reduction.

During the September run (AC\,118S), SOFI suffered a point spread function
variable over the field, while in October the problem was largely solved.

Objects are detected and classified by SExtractor (Bertin \& Arnouts 1996), version
2, using the exposure map for a clean detection.

Because of the shallower images, the 4.4 arcsec aperture (24 Kpc for
cluster galaxies) adopted in Paper I is not an optimal aperture to measure
the flux of our faint galaxies because the aperture integrates mainly
noise (outer regions of faint galaxies are undetected) at a such a large
radius. We adopt, therefore, a smaller (3.0 arcsec) aperture for the flux
determination. By using the deep (central) AC\,118 pointing we verify that
such an aperture misses part of the galaxy flux, even for faint galaxies,
i.e. the 3.0 arcsec aperture magnitude is not a surrogate for the ``total"
magnitude. 

The sample is complete down to $K_s=20$ mag in the most shallow field
(AC\,118S), and therefore the analysis is bound at $K_s\le 20$ mag over
the whole field of view.

\section{Results}

\subsection{Spatial distribution of galaxies of various luminosities}

Figure 1 shows the studied field of view. Each ellipse correspond to one
galaxy\footnote{In the high density regions, the size and orientation of galaxies
are determined with low accuracy because of crowding, but they are never used
in this paper, except in this Figure for pictorial purposes.}.
The three dotted rectangles enclose the three $K_s$
pointings. Attentive inspection of Figure 1 shows that:

-- There are two obvious galaxy overdensities: one in the center and
the other one at 3 arcmin NW.

-- There is another possible overdensity in the far S, at $\sim5$ arcmin 
away from
the center, as can be appreciated by comparing the density of galaxies at
similar distances from the cluster center in the Southern and Northern
pointings. Galaxies in the far S have unknown redshifts, and therefore we don't
know whether this overdensity is associated with the cluster or is a background
group (or cluster). 

\medskip

Figure 2 shows the cluster radial profile, i.e. the number of galaxies per
radius bin, measured in circular annuli. It has been computed for four
magnitude ranges and both including (open points) and excluding (close points)
galaxies in the far S, that are possibly unrelated to the studied cluster, and
in the NW quadrant, where the effect of the NW clump should be higher. The
radial profiles are statistically background subtracted in order to remove
interlopers by using the background galaxy density measured in the HDF-S (da
Costa et al. 2002), as computed by ourselves by using their public images. 
Errors are assumed to be Poissonian (i.e. for the time being we neglect the
intrinsic variance of galaxy counts, that are, instead, taken into account in
the next sections). The area observed at each radius is shown in Figure 3,
and care should be paid to densities computed over a small area
(say, much less than 2  arcmin$^2$) and large clustercentric radii because on these
small areas the intrinsic variance of the
galaxy counts could be very large with respect to the low cluster galaxy
density. 

The cluster radial profile of all galaxies (brighter than $K_s<20$ mag, or 
$M_K \ltapprox -21$ mag, upper--left panel) shows a positive galaxy
density from the center to 2 Mpc away, in particular when galaxies in the
far S are counted. It is centrally peaked. When the NW quadrant is
included in the profile computation, a second broad peak is present at 0.7
Mpc from the cluster center, while at larger radii the profile decreases.
When the quadrant including the NW clump is instead removed, the radial
profile shows a flattening, instead of a second maximum, at $\sim 1$ 
Mpc from the cluster center. 

When we consider only galaxies brighter than $K_s=17$ mag ($M_K\sim-24$ mag),
i.e. massive galaxies, the cluster radial profile (upper--right panel) is
steeper in the center than in the previous case, because the galaxy density
increases by a factor 7 over three bins, to be compared to an increase of a
factor 2 to 3 over the same radial range when all galaxies are considered. The
second maximum is still there when all galaxies are counted (open points). 
Overall, the profile is quite flat outside the cluster core. The evidence of
a positive density at 1.8 Mpc is marginal ($2\sigma$) when discarding 
galaxies in the far S, while it is significant including them.

At the other end of the luminosity function, the radial profile of faint
galaxies, $19<K_s<20$ mag or $-22\ltapprox M_K\ltapprox -21$ mag, is quite flat from
the center to $\sim1-1.2$ Mpc (bottom--right panel), and undetected (i.e.
statistical evidence is $\sim1\sigma$) at large radii, even binning the data
with larger bins.

The shape of spatial distributions of galaxies in the $18<K_s<20$ mag
(bottom--left panel) and $19<K_s<20$ (bottom--righ panel) ranges are quite
similar. There is a factor of two between the amplitudes of the two radial
profiles, because there is a factor of two between the two considered
magnitude ranges, and because the AC\,118 luminosity function is quite
flat at these magnitudes (see Section 3.3). The innermost point seems
higher that the ones at $r\sim1$ arcmin, but without any statistical
significance.

Therefore, the radial profile of all, bright and faint galaxies are quite
different in steepness. Faint galaxies shows similar radial
profiles independent of the two considered magnitude ranges.

\subsection{Dwarfs to giant ratio radial profile}

The left panel of figure 4 shows the giant--to--dwarf ratio, as a function
of the clustercentric distance. For the sake of clarity, galaxies brighter
than $K_s=17$ ($M_K\sim-24$) mag are called giants, while galaxies with
$18<K_s<20$ mag are called dwarfs. The giant--to--dwarf ratio shows a
maximum at the center, where there are similar numbers of giant and dwarfs
in the considered magnitude range, then decreases to a much smaller value
from radii as small as 300 Kpc and as far as 2.2 Mpc. Outside the cluster
core, there are roughly 3 dwarfs per giant in the considered magnitude
ranges. The deficit of dwarfs in the cluster core (or the excess of
giants) is in agreement with that found in the previous Figure and in
Paper I by analysing the shape of the LF at various cluster locations (but
over a restricted cluster portion) and of the giant--to--dwarf ratio at a
few cluster locations. The inclusion or exclusion of the NW quadrant or of
the far S region does not appreciably change the giant--to--dwarf ratio,
as shown in the Figure. The new data presented in this paper do not make
stronger the statistical significance of the found segregation for
clustercentric distance less than 1 Mpc (that it is claimed significant at
$>99.9$ \% confidence level in Andreon 2001), because new data are at
larger clustercentric distances. Since we divide the $R<1$ Mpc range in
three bins, instead of the two bins as in Paper I, the statistical
evidence {\it per bin} is in fact smaller here than in Paper I ($>\sim90$
vs $>99.9$ \% confidence level). At the large clustercentric radii sampled
by the new data, the giant--to--dwarf ratio differs from the central one
at the 90 \% confidence level when all the field of view is considered,
and at the 80 \% confidence level when the NW quadrant and the far S
regions are excluded. In this specific calculation, we take into account
the field--to--field background variance as described in Huang et al.
(1997), and we propagate the errors as described in Gehrels (1986), i.e.
we do not make the simplifying assumption of Gaussian errors.

Similar conclusions can be drawn defining as dwarfs $19<K_s<20$ mag
galaxies (right panel of Figure 4), except that the absolute value of the
giant--to--dwarf ratio increases by approximatively a factor of two,
because the considered magnitude range for dwarfs is now half the size.
The shapes of the giant--to--dwarf radial profiles in the two panel of
Figure 4 are striking similar. This similarity implies that $19<K_s<20$
mag dwarfs are not segregated with respect to $18<K_s<20$ mag dwarfs, as
directly seen in the bottom panels of Figure 2. 

Note, however, that the smaller magnitude range adopted in the right panel
of Figure 4 also decreases the number of dwarfs, and therefore increases
the size of error bars. For the same reason the statistical significance 
of a variation of the giant--to--dwarf ratio is also reduced.

\subsection{Luminosity function at various cluster locations}

The LF is computed as the statistical difference between
(crowding--corrected) galaxy counts in the cluster direction and in the
control field direction. We use the HDF-S (da Costa et al. 2002) as
background (control) field, and we fully take into account the
field--to--field galaxy count fluctuations in the error computation (see
Paper I for details). 

We fitted a spline to the background counts and we
use it in place of the observed data points because background galaxy
counts show an outlier point at $K_s=17$ mag when a 3 arcsec aperture is
adopted. 

The present AC\,118 sample consists of 496 members, about as many galaxies
as in Paper I, but are distributed over a larger area and a narrower
magnitude range. The LF has been fitted by a Schechter (1976) function by
taking into account the finite bin width (details are given in Paper I).
Figure 5 shows the LF computed at different cluster locations and the best
fit Schechter (1976) function to the global (i.e. those measured over the
whole field of view) LF, whose $\phi^*$ is scaled by the ratio between the
number of members at each location and in the global LF. The global LF is
shown in panel $a)$. Panels $b)$ and $c)$ present the LF of the main and
secondary clumps, respectively. Their exact boundary definitions are those
of Paper I (for a pictorial view see Fig. 1 there). Panel $d)$ shows the
LF of galaxies inside the central pointing but outside the two clumps.
Panels $e)$ shows the LF of galaxies in the Northern and Southern
pointings (not overlapping with the central pointing) and not in the far
S. Finally, panel $f)$ shows the LF of the galaxies in the far S (southest
1 arcmin). 

The best fit parameters to the global LF are: $K^*_s=16.4$ mag
($M_{K^*_s}\sim-24.9$ mag) and $\alpha=-0.85$, where $\alpha$ is the slope
of the faint part of the LF, and $M^*$ is the knee of the LF, i.e. the
magnitude at which the LF starts to decrease exponentially. We re--state
that the present 3 arcsec magnitude misses a significant part of the
galaxy flux, and hence the found parameters should not be used for, say,
computing the luminosity density, or for comparison with values derived
from other samples using a different metric (or any isophotal) aperture.
This could also be appreciated by noting that in Paper I, using magnitudes
that include a large fraction of the galaxy flux, we found steeper LFs
than shown in panel $b)$, $c)$ and $d)$ for the same considered cluster
and background regions. Here we use the LF as a tool for comparing the
abundance of galaxies of various luminosities in different environments
for a sample of galaxies all at the same redshift and whose flux is
measured in one single way. A thorough discussion of the cosmological
implication of lost flux from galaxies is given in Wright (2001) and
Andreon (2002). Errors, quoting the projection of the $\Delta\chi^2=2.3$
(68\% for two interesting parameters, Avni 1976) confidence contours on
the axis of measure are: 0.35 mag and 0.21, respectively, for $K^*_s$ and
$\alpha$. The conditional errors, i.e. the errors when the other
parameters are kept at the best values (that has a low statistical sense,
Press et al. 1992) are found to be at least half the size\footnote{and are
often quoted in the literature (forgetting the adjective ``conditional").} .
The AC\,118 global LF is smooth and is well described by a Schechter function
($\chi^2/\nu\sim4.2/8$). 

The parameters of the global LF also describe the shape of the LF measured at
other cluster locations (see panels from $c)$ to $f)$), because the reduced
$\chi^2$ is of the order of 1 or less, except for the LF in panel $b)$.
Galaxies considered in panel $b)$ are in the cluster center: for the total
number of observed galaxies there are a too many very bright galaxies (say,
brighter than $K_s=16-17$ mag) and too few fainter galaxies, an effect already
found in Paper I for the same region and using the same data, but adopting a
magnitude definition which includes a larger galaxy flux. This is the same
effect shown in Figures 2 and 4 and presented in the previous sections,
measured here by looking for differences in the LF computed at several cluster
locations instead of looking for a dependence between the spatial distribution
of galaxies and their luminosities. Differences found in Paper I are confirmed
here (by adopting a $\sim95$ \% confidence level threshold and using a
Kolmogorov--Smirnov test, that is preferable to comparing the best fit values
because of the correlation between parameters  and of the need for an assumption
of a given parental distribution): the LF is flatter at the main clump (panel
$b)$) than at all the other considered regions. All the other LFs are
compatible each other at better than 95\% confidence level, extending at
larger radii the findings in Paper I: the LF steepens going from high-- to low--
density environments and the steepening stops in the region considered in panel
$d)$. The new result is that the LF does not change in regions not surveyed in
Paper I, i.e. for galaxies whose average clustercentric projected distance is
1.2 Mpc (for galaxies in the N and S pointings, panel $e)$) and 1.8 Mpc (for
galaxies in the far S, panel $f)$.

The $f)$ panel only includes galaxies in the far S (southest 1 arcmin).
These galaxies are an extension of the AC\,118 cluster, or
another group (or part of a cluster) along the line of sight. 
Given the small number of galaxies in this region (56 galaxies out of 535)
and the similarity of their LF to the global one, their inclusion or
exclusion from the global LF makes no difference.

The LFs computed thus far can be used to test whether the galaxy overdensity in
the far S is at the AC\,118 redshift, under the assumption that the LF is a
standard candle outside the cluster core. The use of the near--infrared LF as a
standard candle has been exploited by de Propris et al. (1999) to
study the luminosity evolution of galaxies up to $z\sim1$.  There are two paths
for the computation, depending on whether a parametric form is used for the LF
shape (and in such a case the errors on the data points are included in the
confidence level calculation) or no (that neglects
errors on data points). For galaxies in the far S sample, the 68 \% conditional
confidence range (i.e. once $\alpha$ is keep fix to the best fit value) for
$M^*$ are 16.3 and 17.9 mag, limiting the difference in distance modulus
between AC\,118 and the far S overdensity to $\Delta (m-M)= -0.1,+1.5$ mag or,
in redshift, -0.01,+0.3. This range in $\Delta (m-M)$ only excludes that the
galaxies in the far S are in the AC\,118 foreground. To be precise, 
the high redshift constraint is broader, because our 3 arcsec aperture
includes more and more galaxy flux as the redshift increases, and we have not
accounted for this effect.  By using the data points alone and a
Kolmogorov--Smirnov test, the 68 \% confidence range is $\Delta
(m-M)=0,\sim1.7$ mag, quite similar to the parametric result. Therefore, the
analysis of the LF is not sufficient to say whether these galaxies belong to
the cluster of are in the background of AC\,118. Surely, these galaxies 
do not lie in front of the cluster.

In conclusion, the analysis of the luminosity function shows the same
luminosity segregation found in the analysis of the galaxy spatial
distribution. With respect to Paper I, we extended the analysis to much larger
distances (1.8 Mpc vs 0.58 Mpc).

\section{Summary and discussion}

We detect AC\,118 from the center to half the Abell radius (1.5 Mpc) and
possibly to 2.0 Mpc. 

There is a luminosity segregation among the galaxies in the AC\,118
cluster and it has been shown in three different ways: by studying the LF
dependence on environment, by a radial analysis of the dwarf--to--giant
ratio and by comparing the radial profiles of galaxies of different
luminosities. While the three methods differ, they are not independent. It
is the order in which the grouping is done that changes: galaxies are
first grouped spatially and then their luminosity distribution is studied
in the LF analysis, while in the two other methods galaxies are first
grouped in luminosity and then their spatial distribution is studied. 

Any choice of cosmology rigidly moves the upper abscissa of Figure 2, 3, 4 and 5 by
a fixed amount, and does not change the shape or relative 
differences of the plotted profiles. Therefore, the detection of a luminosity
segregation in AC\,118 is independent of the choice of the cosmological
values.

The segregation concerns mainly the inner 250 Kpc of the cluster (see in
particular Figure 4), while at larger radii all galaxies have the same
spatial distribution regardless of the galaxy near--infrared luminosity,
up to 2 Mpc away from the cluster center. The segregation consists of an
excess, of a factor of 3, of giants galaxies in the cluster core (or in a
deficit, of the same factor, of dwarf galaxies). Since the numerical
density of dwarfs is largely constant, the luminosity segregation found
seems due to an excess (relative to the number of dwarf galaxies) of giant
galaxies in the cluster center, and not due to a deficit of dwarfs in the
remaining of the cluster. 

Beside AC\,118, luminosity segregation in the near--infrared has been suggested
in the Coma cluster (Andreon \& Pell\'o 2000), although through comparison of
heterogeneous data.

The luminosity segregation found in Paper I is here confirmed to hold over 
a even
wider cluster region.  With respect to the previous investigation on AC\,118,
we take two more paths for confirming the luminosity segregations: the analysis
of the galaxy spatial distribution, and the computation of the radial profile
of the giant to dwarf ratio. Our results are in broad agreement with what 
has been found
in similar analyses, but performed at optical wavelengths (Zwicky 1957, Mellier
et al. 1988; Driver, Couch \& Phillipps 1998; Secker, Harris \& Plummer 1997;
Garilli, Maccagni \& Andreon 1999), or by using the velocity segregation
(Chincarini \& Rood 1977; Struble 1979; Biviano et al. 1992; Stein 1997), or by
analysing the galaxy angular correlation function (Loveday et al. 1995): these
studies found that brightest galaxies are more tightly correlated (or have
lower velocity dispersions) than the faintest galaxies.

Therefore, there is clear evidence of luminosity segregation. Since the
near--infrared luminosity is a good tracer of the stellar mass (Bruzual \&
Charlot 1993), the segregation found is interpreted a mass--related
segregation. The luminosity segregation we found in the near--infrared 
implies a mass segregation more tightly that under the usual assumption
than optical luminosity traces mass: here we show directly that massive
galaxies are found preferentially in the cluster center.

A mass--related segregation is a natural expected outcome of a hierarchical
scenario of cluster formation, because the clustering strength depends on the
halo circular velocity (and therefore on mass) in cold dark matter models (White
et al. 1987, Kauffmann et al. 1997).  However, the effect has been detected only
recently in the simulations (Springer, White, Tormen \& Kauffman 2001) and a
quantitative comparison between observations and simulations awaits a prediction
in a more suitable form. 

The hostile cluster environment plays a role in shaping
the AC\,118 LF but only at small clustercentric radii (or high density),
since outside the cluster core the LF computed at several locations
are all compatible with each other and the dwarf--to--giant ratio is constant
within the errors.

\begin{acknowledgements}
This work is part of a collaboration with M. Arnaboldi, G. Busarello, M.
Capaccioli, G. Longo, P. Merluzzi and G. Theureau. A. Wolter and A. Iovino are
acknowledged for useful discussions. The near--infrared observations presented
in this paper have been taken during the NTT guaranteed time of Osservatorio
di Capodimonte. The director of my institute, Prof. M. Capaccioli, is warmly
thanked for permitting me a long stay at the Osservatorio Astronomico di
Brera, where this work has been prepared. The director of the latter institute,
Prof. G. Chincarini, is acknowledged for hospitality. Comments from the referee
helped to improve the presentation of this paper.
\end{acknowledgements}


\newpage

\begin{table*}
\caption{The data}
\begin{tabular}{lcccc}
\hline
 & AC\,118C & AC\,118S & AC\,118N \\
\hline
\\
Exposure time (min) & 265 & 75 & 90.7 \\
Seeing (FWHM, arcsec) & 0.75 & 1.0 & 1.2 \\
Fully corrected noise$^a$ (mag arcsec$^{-2})$ & 24.0 & 23.0 & 23.3 \\
Applied Illumination correction? & no & yes & no \\
Photom. zero--point RMS over the field (mag) & 0.007 & 0.013 & 0.004 \\
\\
\hline
\end{tabular}
\null$^a$ The sky noise is measured as the dispersion of
adjacent pixels in the background, once the image is binned 
in pixels of 1 arcsec.
\end{table*}
\newpage
\null
\newpage

\begin{figure*}
\psfig{figure=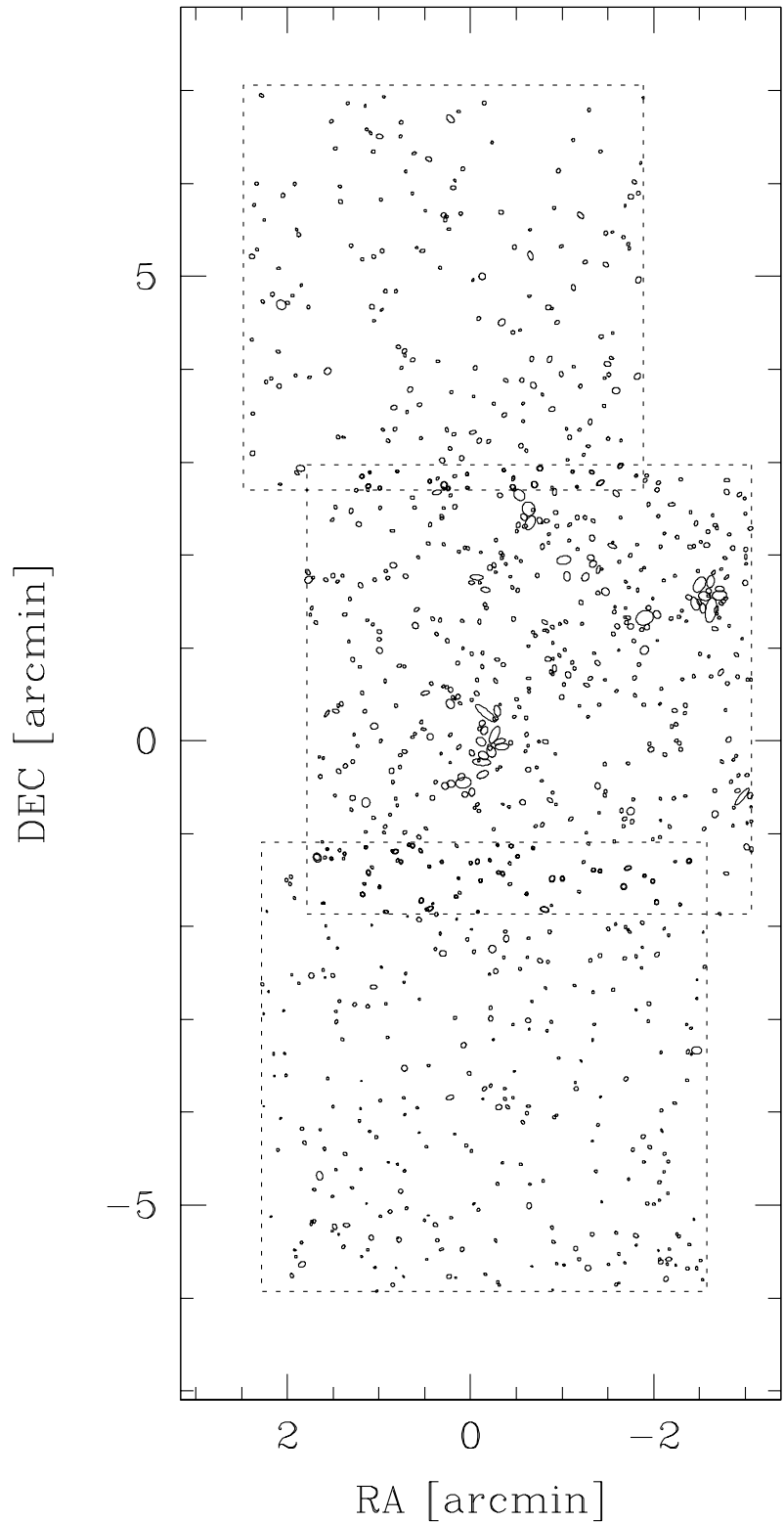,height=21truecm}
\caption[h]{Detected galaxies (ellipses), brighter than $K_s=20$ mag, and X--ray
contours. The area of ellipses is twice the galaxy isophotal area computed at 
$1.5\sigma$ above the background (20.8-21.8 mag arcsec$^{-2}$). 
North is up and East
is to the left. The origin is at ($\alpha, \delta) = (00 \ 14 \ 20, \ -30 \ 24 
\ 00)$ (J2000)}
\end{figure*}

\newpage

\begin{figure*}
\hbox{\psfig{figure=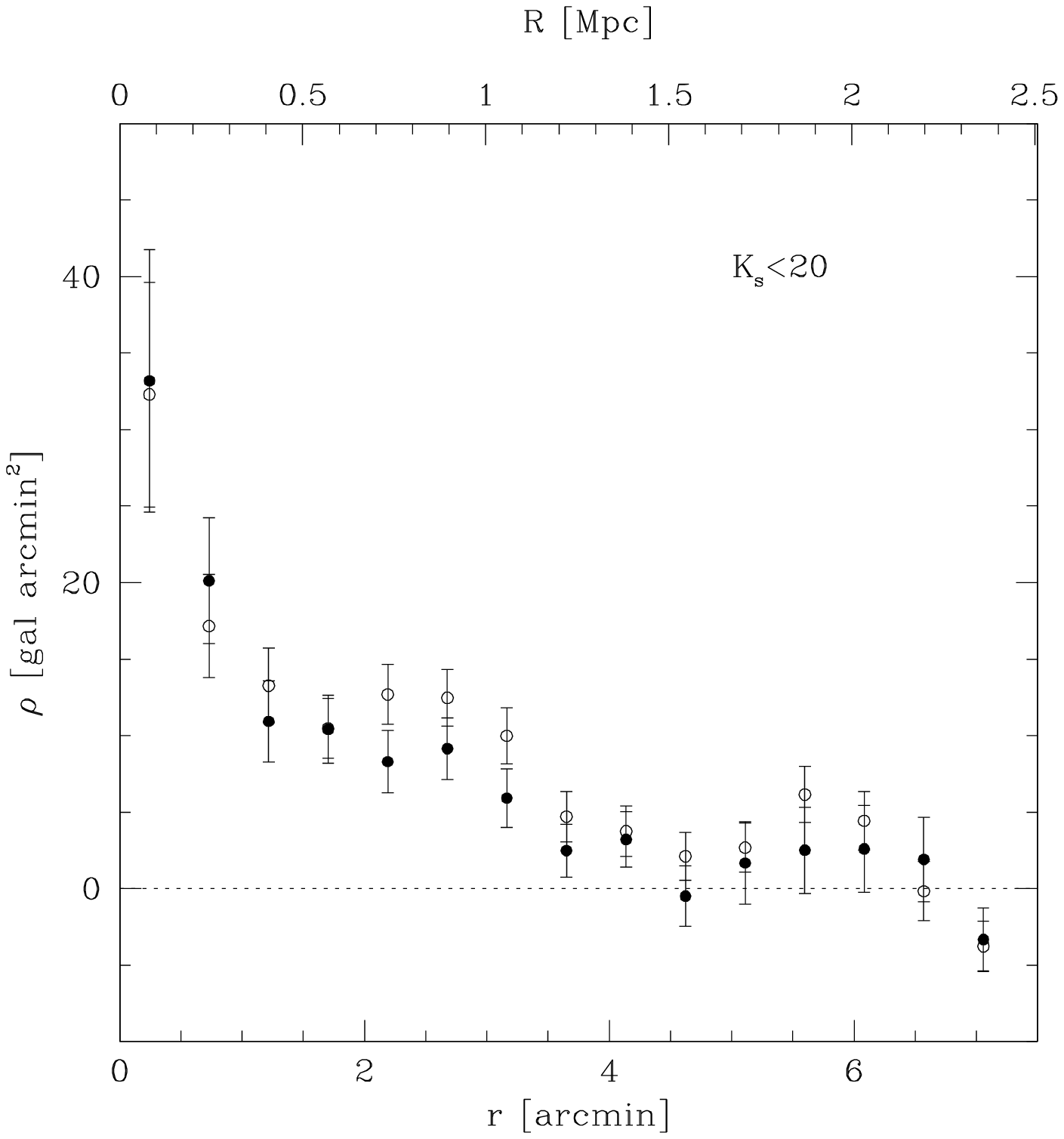,height=8truecm}\psfig{figure=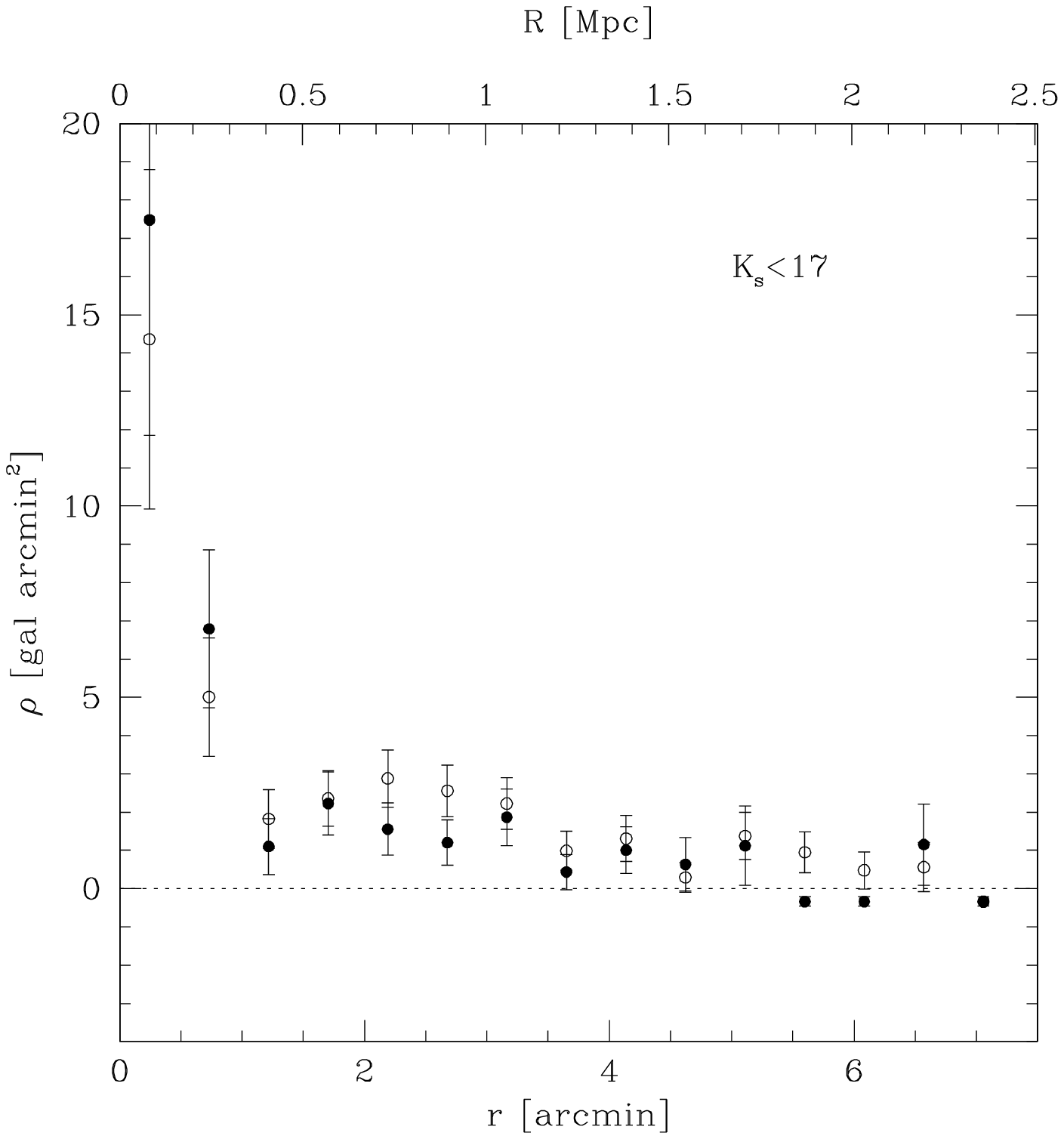,height=8truecm}}
\hbox{\psfig{figure=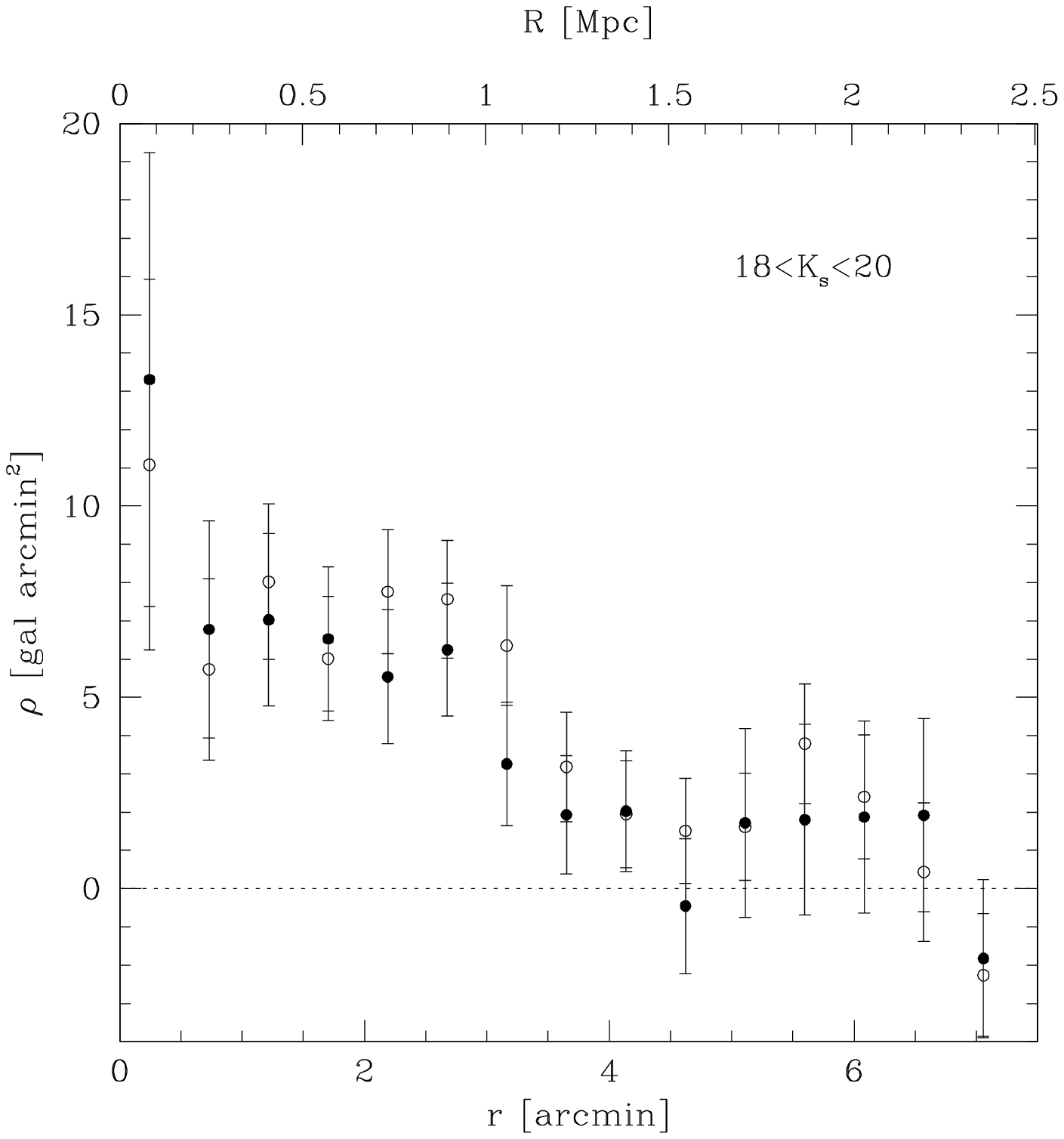,height=8truecm}\psfig{figure=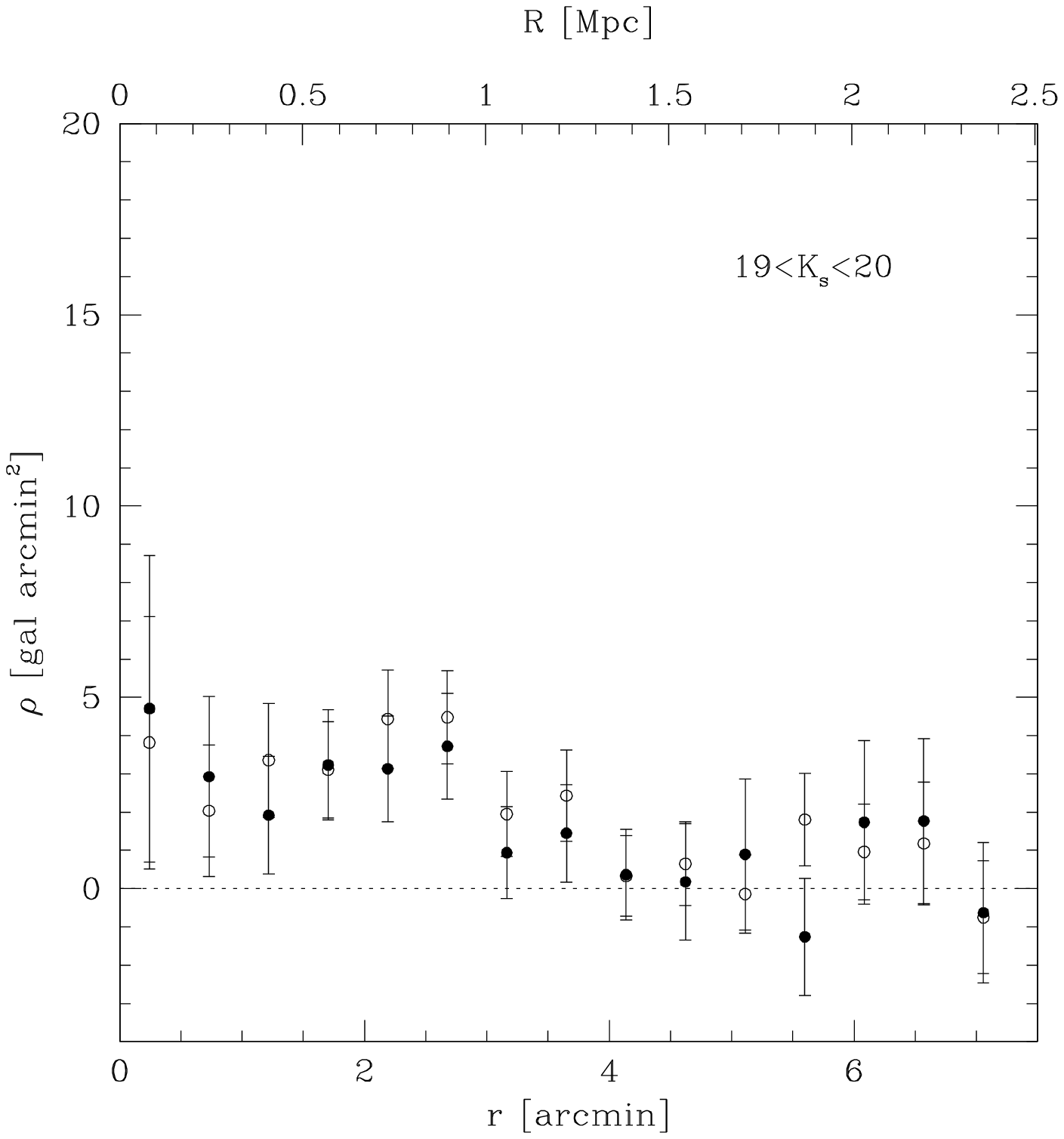,height=8truecm}}
\caption[h]{Background subtracted radial profiles 
of the AC\,118 cluster. Galaxies having
$K_s<20$, $K_s<17$, $18<K_s<20$ and $19<K_s<20$ mag are selected for the computation of the
radial profiles in the top--left, top-right, bottom--left and bottom--righ panels,
respectively. Open dots show the profile computed over all the observed field, 
while closed points show the profile computed excluding
the southest 1 armin and the NW quadrant. Note the steepness of the radial profile of
bright galaxies (upper--right panel) and the flatness of the radial
distribution of faint galaxies (lower--left panel). The density
of background galaxies measured in the HDF-S is 7.8, 0.3, 6.3, 4.2 
gal arcmin$^{-2}$ for $K_s<20$, $K_s<17$, $18<K_s<20$
and $19<K_s<20$ mag respectively, when
adopting our 3 arcsec aperture magnitude, and has been already subtracted.} 
\end{figure*}

\begin{figure*}
\psfig{figure=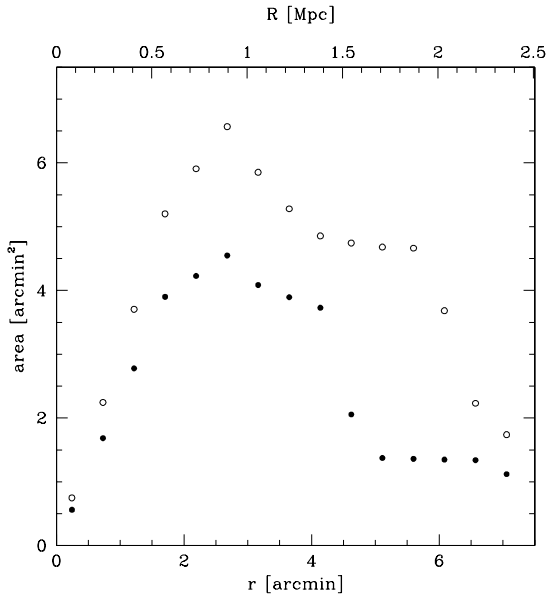,height=8truecm}
\caption[h]{Area over which the radial profiles are computed. Open points mark
the area studied when all the field of view is considered, while close points 
mark the area when the NW quadrant and the southest 1 armin are excised.}
\end{figure*}

\begin{figure*}
\hbox{\psfig{figure=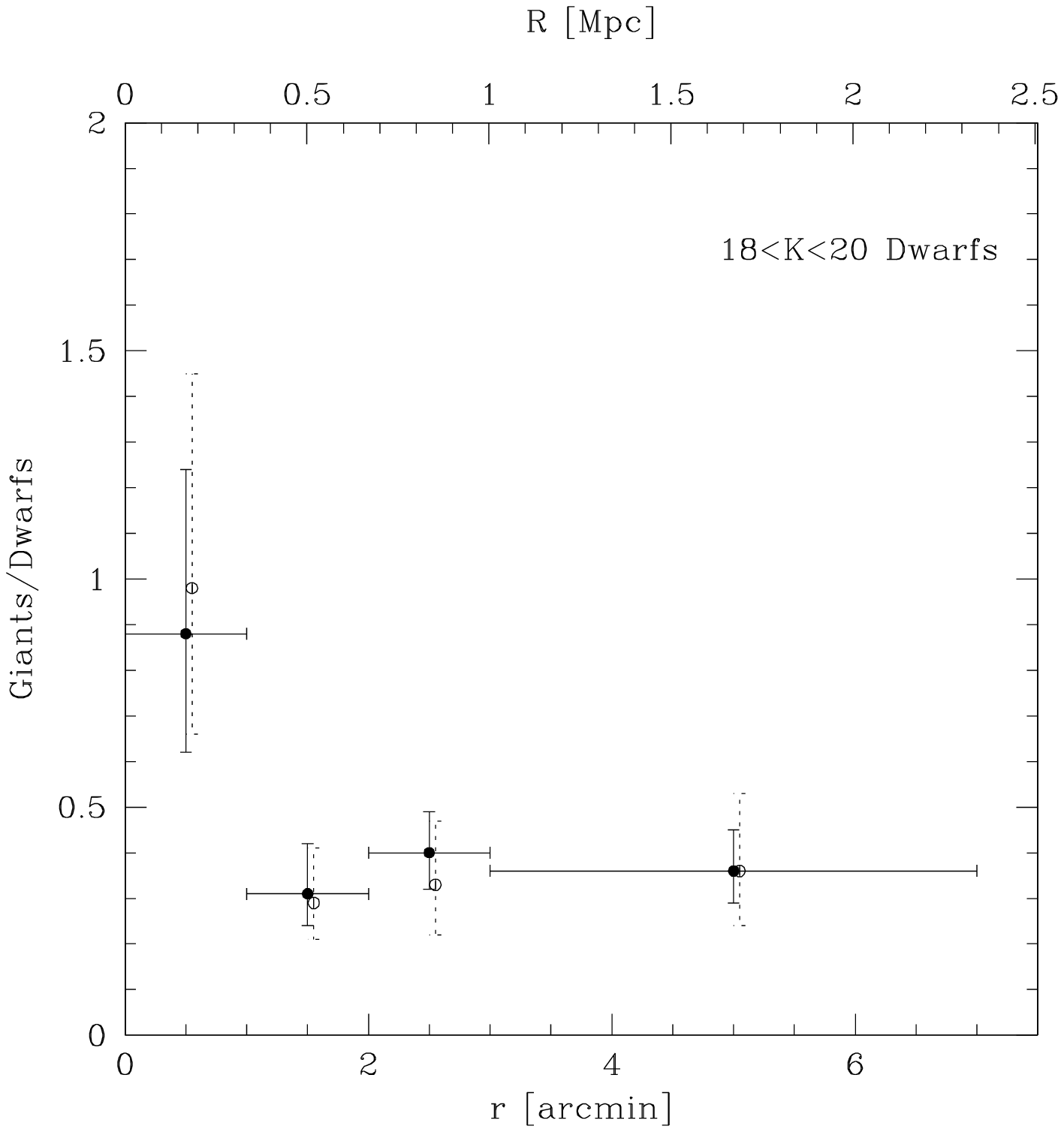,height=8truecm}\psfig{figure=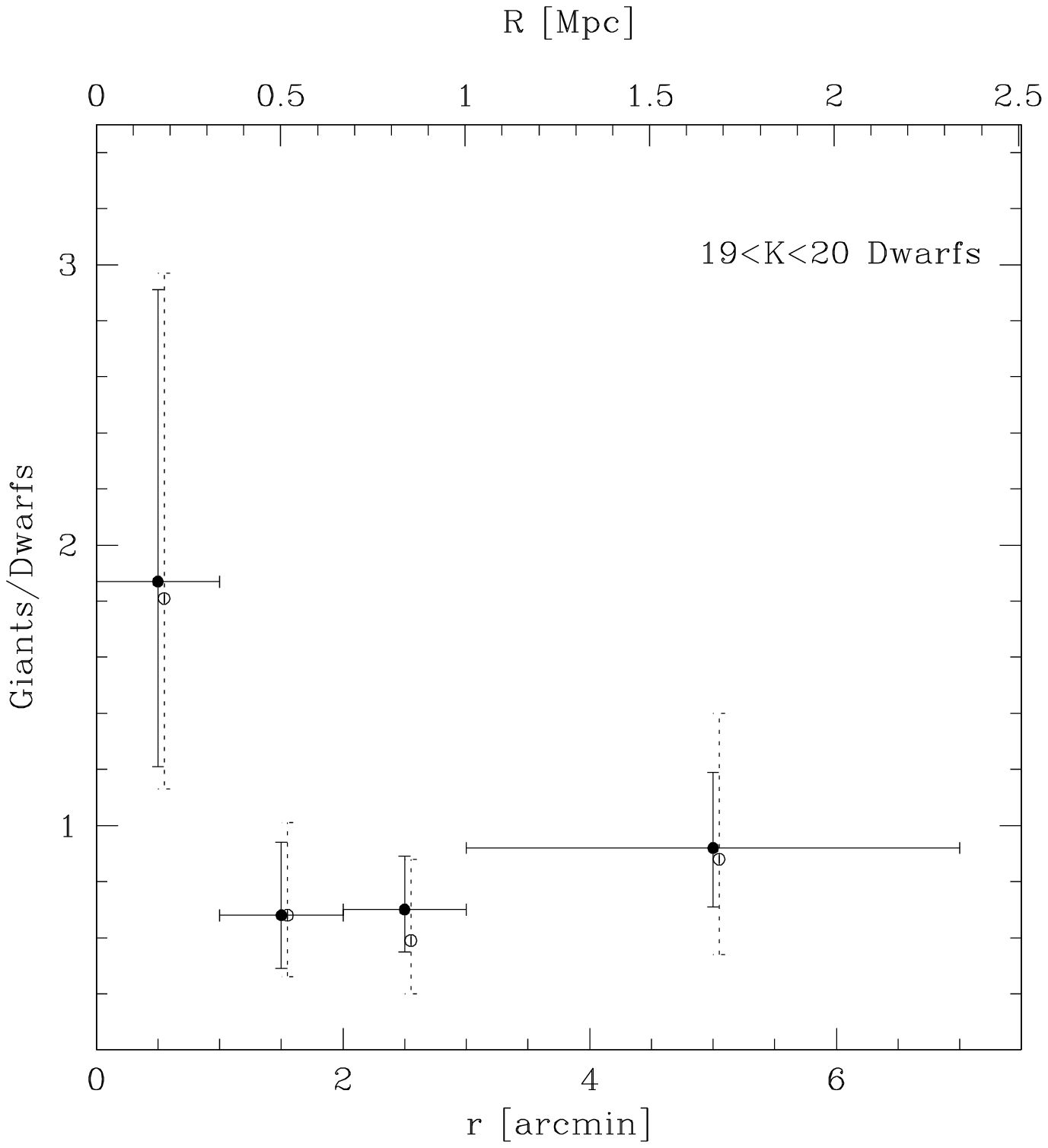,height=8truecm}}
\caption[h]{Giant ($K_s<17$) to dwarf ($18<K_s<20$ in the left panel, 
$19<K_s<20$ in the right panel)
ratio as a function of the clustercentric distance, including and excluding the NW quadrant and the
southest 1 armin (open and solid points, respectively). Error bars in the abscissa show the bin width.
For display purposes error bars in the abscissa are drawn once and points are slightly displaced in x.}
\end{figure*}

\begin{figure*}
\psfig{figure=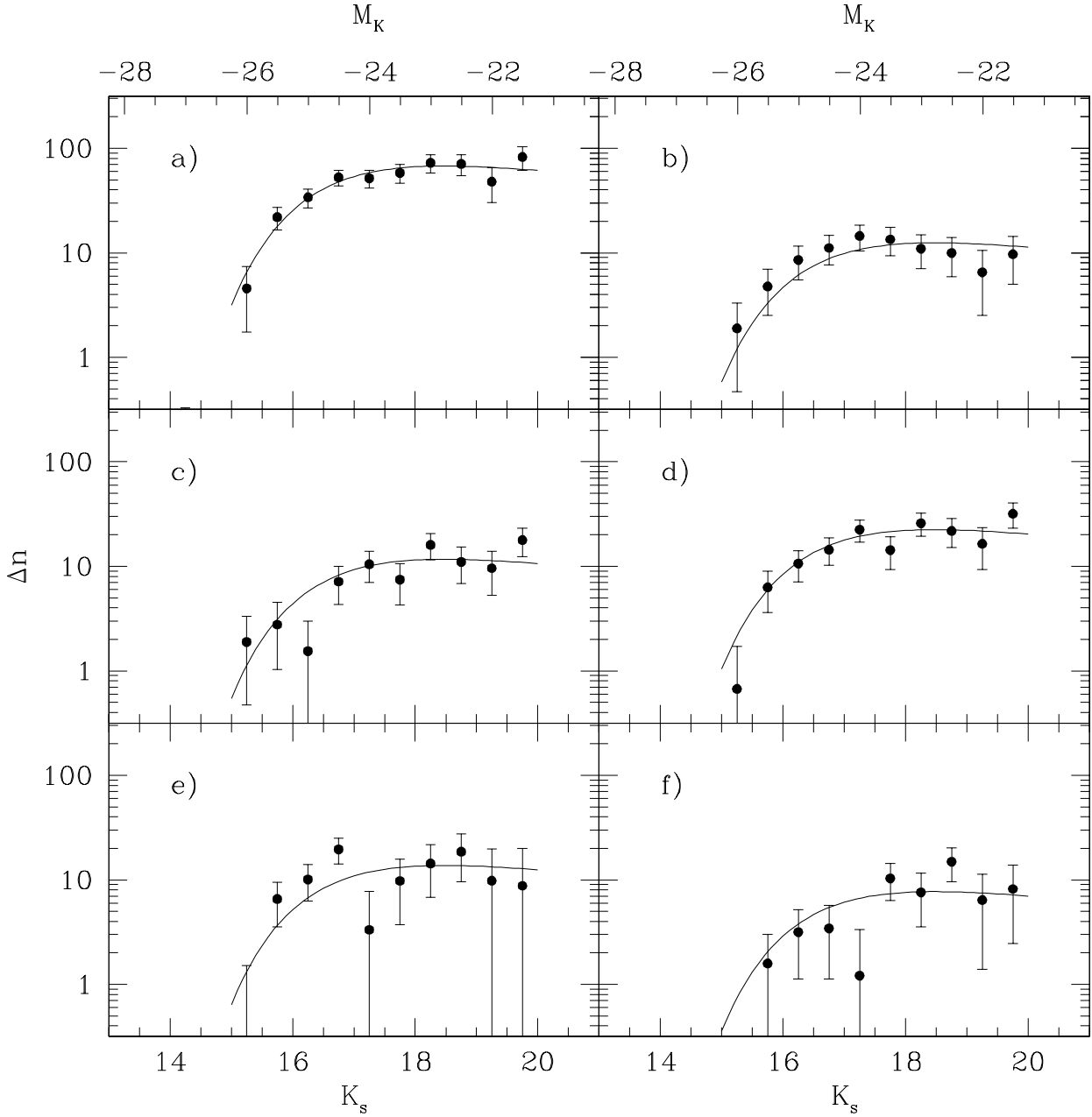,height=17truecm}
\caption[h]{Luminosity function of AC\,118. The global (i.e. integrated over
the whole studied field) LF is shown in panel $a)$. Panels $b)$ and $c)$
present the LF of the main and secondary clumps, respectively. Panel $d)$
is the LF of galaxies inside the central pointing but outside the two clumps.
Panels $e)$ shows the LF of galaxies in the Northern and Southern pointings
(not overlapping to the central pointing) and not in the far S. Finally,
panel $f)$ shows the LF of the galaxies in the far S (southern 1 arcmin). 
The curve is the best fit function to the whole cluster, with $\phi^*$ 
adjusted to reproduce the total number of galaxies at each considered 
location. There are 496, 91, 86, 164, 101, 56 galaxies in panel a), b), c), d)
e) and f), respectively.}
\end{figure*}

\end{document}